\documentclass[12pt]{article}
\usepackage{amsmath,amsfonts,amsbsy}
\usepackage{pstricks,pst-node}

\psset{unit=1.3cm,linewidth=.5pt,radius=.2}  

\usepackage{enumitem}                       
\usepackage{multirow}                     
\usepackage{float}                          
\usepackage{lscape}                         
\usepackage{bm}

\def\hybrid{\topmargin -20pt    \oddsidemargin 0pt
        \headheight 0pt \headsep 0pt
        \textwidth 6.25in       
        \textheight 9.5in       
        \marginparwidth .875in
        \parskip 5pt plus 1pt   \jot = 1.5ex}

\hybrid

\newcommand{\beq}{\begin{equation}}
\newcommand{\eeq}{\end{equation}}
\newcommand{\bi}{\begin{itemize}}
\newcommand{\ei}{\end{itemize}}
\newcommand{\bt}{\begin{tabular}}
\newcommand{\et}{\end{tabular}}
\newcommand{\bc}{\begin{center}}
\newcommand{\ec}{\end{center}}

\newcommand{\g}{\mathfrak{g}}

\newcommand{\be}{\begin{equation}}
\newcommand{\ee}{\end{equation}}
\newcommand{\bea}{\begin{eqnarray}}
\newcommand{\eea}{\end{eqnarray}}
\newcommand{\ba}{\begin{array}}
\newcommand{\ea}{\end{array}}

\def\bbox{{\,\lower0.9pt\vbox{\hrule \hbox{\vrule height 0.2 cm
\hskip 0.2 cm \vrule height 0.2 cm}\hrule}\,}}
\newcommand{\dsl}{\pa \kern-0.5em /}
\renewcommand{\t}{\theta}

\def\a{\alpha}
\def\b{\beta}
\def\g{\gamma}
\def\G{\Gamma}
\def\d{\delta}
\def\e{\epsilon}

\def\l{\lambda}
\def\m{\mu}
\def\n{\nu}

\def\r{\rho}
\def\s{\sigma}

\def\t{\tau}

\def\de{\partial}

\makeatletter \@addtoreset{equation}{section} \makeatother

\def\slashchar#1{\setbox0=\hbox{$#1$}           
   \dimen0=\wd0                                 
   \setbox1=\hbox{/} \dimen1=\wd1               
   \ifdim\dimen0>\dimen1                        
      \rlap{\hbox to \dimen0{\hfil/\hfil}}      
      #1                                        
   \else                                        
      \rlap{\hbox to \dimen1{\hfil$#1$\hfil}}   
      /                                         
   \fi}

\begin{document}

\begin{titlepage}
\begin{center}

\hfill UG-10-21 \\
\hfill IFT-UAM/CSIC-10-17 \\
\hfill KCL-MTH-10-03 \\

\vskip 1cm

{\Large \bf  IIA/IIB Supergravity and Ten-forms}

\vskip 1.5cm

{\bf E.A.~Bergshoeff\,$^1$, J. Hartong\,$^2$, P.S. Howe\,$^3$, T.
Ort\'\i n\,$^4$ and F. Riccioni\,$^3$}

\vskip 30pt

{\em $^1$ \hskip -.1truecm Centre for Theoretical Physics, University of Groningen, \\
Nijenborgh 4, 9747 AG Groningen, The Netherlands \vskip 5pt }

{email: {\tt E.A.Bergshoeff@rug.nl}} \\

\vskip 15pt

{\em $^2$ \hskip -.1truecm Albert Einstein Center for Fundamental Physics,
Institute for Theoretical Physics,
University of Bern,
Sidlerstrasse 5, 3012 Bern, Switzerland\vskip 5pt }

{email: {\tt  hartong@itp.unibe.ch}}

\vskip 15pt

{\em $^3$ \hskip -.1truecm Department of Mathematics King's College
London Strand London WC2R 2LS UK}

{email: {\tt Paul.Howe,Fabio.Riccioni@kcl.ac.uk}}

\vskip 15pt

{\em $^4$ \hskip -.1truecm Instituto de F\'isica Te\'orica UAM/CSIC
Facultad de Ciencias C-XVI, C.U. Cantoblanco, E-28049-Madrid, Spain}

{email: {\tt Tomas.Ortin@uam.es}}

\end{center}

\vskip 1cm

\begin{center} {\bf ABSTRACT}\\[3ex]
\end{center}

We perform a careful investigation of which $p$-form fields can be
introduced consistently with the supersymmetry algebra of IIA and/or
IIB ten-dimensional supergravity. In particular the ten-forms, also
known as ``top-forms'', require a careful analysis since in this
case, as we will show, closure of the supersymmetry algebra at the
linear level does {\emph not} imply closure at the non-linear level.
Consequently, some of the (IIA and IIB) ten-form potentials
introduced in earlier work of some of us are discarded. At the same
time we show that new ten-form potentials, consistent with the full
non-linear supersymmetry algebra can be introduced. We give a
superspace explanation of our work. All of our results are precisely
in line with the predictions of the $E_{11}$ algebra.

\begin{minipage}{13cm}
\small

\end{minipage}




\vfill


\end{titlepage}

\newpage
\setcounter{page}{1}

\newpage


\section{Introduction}\setcounter{equation}{0}

Supergravity theories provide important information about string
theory. In particular, the $p$-form fields of the supergravity
multiplet are in one-to-one correspondence, via their occurrence in
the world-volume actions, with the branes of string theory provided
that supersymmetry can be maintained.\footnote{An exception are
those $p$-form fields that under supersymmetry do not transform into
the gravitino. An example of this is the 8-form dual of the IIA
dilaton, see the last line of eq.~\eqref{purelyfermionicIIA}.} The
$p$-form fields with $p\le D-2$ can be easily predicted since they
describe the physical states (or their duals) of the (D-dimensional)
supergravity theory in question. This is not the case for the
potentials of rank $p=D-1$ (``de-form'' potentials) and rank $p=D$
(``top-form'' potentials). A well-known example of a de-form
potential is the 9-form potential \cite{Bergshoeff:1996ui} of
massive IIA supergravity \cite{Romans:1985tz} that gives a dual
description of the mass parameter $m$ present in the theory.

Ten-dimensional supergravities have been constructed a long time ago
both for the non-chiral IIA case
\cite{Giani:1984wc,Campbell:1984zc,Huq:1983im} as well as for the
chiral IIB case \cite{Schwarz:1983wa,Schwarz:1983qr,Howe:1983sra}. A
partially ``democratic'' formulation of these theories, where all
the RR fields are introduced together with their magnetic duals, was
given in \cite{Bergshoeff:2001pv} and \cite{Bergshoeff:1999bx}.
This was then extended to a fully democratic
formulation, including also the magnetic duals of the NS fields for
both the IIA \cite{Bergshoeff:2006qw} and IIB
\cite{Bergshoeff:2005ac} cases. This analysis also included a fairly
complete discussion of the de-form and top-form potentials for IIA
and IIB. A superspace interpretation of the latter case was given in
\cite{Bergshoeff:2007ma}.

It is the purpose of this note to reconsider the results of
\cite{Bergshoeff:2006qw,Bergshoeff:2005ac,Bergshoeff:2007ma}. The
reason for this is the following. In \cite{Bergshoeff:2005ac} we
showed that an ${\rm SU(1,1)}$-doublet of ten-form potentials could
be introduced, consistently with the lowest-order IIB supersymmetry
algebra, with only trivial Abelian gauge transformations:
\begin{equation} \label{trivial}
\delta A_{10}^\alpha = d \Lambda_9^\alpha\,,\hskip 2truecm
\alpha=1,2.
\end{equation}
In contrast, both $E_{11}$ \cite{Kleinschmidt:2003mf} and IIB
superspace \cite{Bergshoeff:2007ma} considerations predict a doublet
ten-form potential that forms a  non-trivial gauge algebra with the
other $p$-forms of the theory. Moreover, as we will show in this
paper, IIB superspace does not allow for \emph{two} doublets. We
will show that the different inconsistencies are resolved as
follows. By performing an explicit check of the full non-linear
supersymmetry algebra, we show that, whereas at the lowest-order
level two doublets of ten-form potentials are allowed, this is no
longer the case at the non-linear level: surprisingly, we find that
the trivial doublet \eqref{trivial} is \emph{not} consistent with
the full IIB supersymmetry algebra. The same result applies to the
IIA case. This is a rare example of a case that a result,
established at the lowest-order level of the supersymmetry algebra,
cannot be extended to the full non-linear level. It relies on the
fact that top-forms are special in the sense that a general
coordinate transformation can be rewritten as a gauge
transformation. Consequently, when closing the algebra at the lowest
order in fermions one ``only'' has to make sure that the algebra
closes up to  gauge transformations while for the lower rank
potentials one needs both gauge transformations and g.c.t.'s.

On the other hand, IIB superspace and $E_{11}$ considerations do
predict the existence of a doublet of ten-form potentials with a
non-trivial gauge-symmetry structure which were not found in
\cite{Bergshoeff:2005ac}. We show that this doublet was missed
because of a specific identity that was not used in the
calculations. Here, we point out this identity and show that the
supersymmetry algebra can now indeed be closed. For the IIA case we
establish a similar result: we show that the trivial ten-form found
in \cite{Bergshoeff:2006qw} does not persist at the non-linear
level. Instead, by using a similar non-trivial identity as in the
IIB case, we show that a new ten-form potential is allowed, with a
non-trivial gauge algebra structure, that is consistent  with the
full non-linear IIA supersymmetry algebra.

This paper is organised as follows. In section \ref{IIB} we first
discuss the ten-form potentials of IIB supergravity. In section
\ref{IIA} we give a similar discussion of the IIA case. Finally, in
section \ref{conclusions} we give our conclusions.

\section{The Top-forms of IIB Supergravity} \label{IIB}
In this section we want to reconsider the analysis of ten-form
potentials present in the IIB supergravity multiplet that was
performed in \cite{Bergshoeff:2005ac}. We will first review the
supersymmetry algebra and the algebra of gauge transformations for
all the propagating fields. We will then consider in more detail
the 10-forms. Finally, we will review the superspace results of
\cite{Bergshoeff:2007ma}.

\subsection{IIB supergravity}
The propagating fields of IIB supergravity and their magnetic duals
are the vielbein $e_\m{}^a$, two scalars parametrising the symmetric
manifold $SU(1,1)/U(1)$ and described in terms of the matrix
$V^{\alpha}_\pm$, where $\pm$ denotes the $U(1)$ charge and $\a$ is
a doublet index of $SU(1,1)$, a doublet of 2-forms $A^\a_{\m_1
\m_2}$ and a self-dual 4-form $A_{\m_1 ...\m_4}$ that is a singlet
of $SU(1,1)$, together with a doublet of 6-forms $A^\a_{\m_1
...\m_6} $ and a triplet of 8-forms $A^{\a\b}_{\m_1 ...\m_8}$. The
gauge transformations of the form fields can be written in an
abelian basis in which all gauge transformations commute, and in
particular one gets
  \begin{eqnarray}
  \d A^\a_{\m_1 \m_2 } &=& 2 \de_{[\m_1 } \Lambda^\a_{\m_2 ]}\quad\nonumber ,\\
  \d A_{\m_1 ...\m_4} &=& 4 \de_{[\m_1} \Lambda_{\m_2 ...\m_4 ]} - \tfrac{i}{4} \e_{\g\d}
  \Lambda^\g_{[\m_1}F^\d_{\m_2 ...\m_4 ]}
  \quad\nonumber ,\\
  \d A^\a_{\m_1 ...\m_6} &=& 6 \de_{[\m_1} \Lambda^\a_{\m_2 ...\m_6]} - 8 \Lambda^\a_{[\m_1}
  F_{\m_2...\m_6 ]} - \tfrac{160}{3}
  F^\a_{[\m_1 ...\m_3}\Lambda_{\m_4...\m_6]}\quad\nonumber ,\\
  \d A^{\a\b}_{\m_1 ...\m_8} &=& 8 \de_{[\m_1} \Lambda^{(\a\b)}_{\m_2 ...\m_8 ]} +\tfrac{1}{2}
  F^{(\a}_{[\m_1 ...\m_7 } \Lambda^{\b)}_{\m_8 ]}
  - \tfrac{21}{2} F^{(\a}_{[\m_1 ...\m_3 } \Lambda^{\b)}_{\m_4 ...\m_8 ]} \quad ,
  \label{bosgaugeIIB}
  \end{eqnarray}
where the corresponding gauge invariant field strengths are
\begin{eqnarray}
F^\a_{\m_1 ...\m_3} &=& 3 \de_{[\m_1} A^\a_{\m_2 \m_3 ]} \quad\nonumber ,\\
F_{\m_1 ...\m_5} &=& 5 \de_{[\m_1} A_{\m_2 ...\m_5 ]} +
\tfrac{5i}{8} \e_{\a\b} A^\a_{[\m_1 \m_2} F^{\b}_{\m_3 ...\m_5 ]}
\quad\nonumber ,\\
F^{\a}_{\m_1 ...\m_7} &=& 7 \de_{[\m_1 }  A^\a_{\m_2 ...\m_7 ]} + 28
A^\a_{[\m_1 \m_2} F_{\m_3 ...\m_7 ]}
  - \tfrac{280}{3} F^\a_{[ \m_1 ...\m_3 }A_{\m_4 ...\m_7 ]} \quad\nonumber ,\\
F^{\a\b}_{\m_1 ...\m_9} &=& 9 \de_{[\m_1} A^{\a\b}_{\m_2 ...\m_9 ]}
+ \tfrac{9}{4} F^{(\a}_{[\m_1 ...\m_7} A^{\b)}_{\m_8\m_9 ]}
 - \tfrac{63}{4} F^{(\a}_{[\m_1 ...\m_3}A^{\b)}_{\m_4 ...\m_9 ]}  \quad
 \label{IIBfieldstrengths}.
\end{eqnarray}
The supersymmetry transformations of these gauge fields
   \begin{eqnarray}
   \d A^\a_{\m_1 \m_2} &=& \d_F  A^\a_{\m_1 \m_2}\quad,
\nonumber\\
   \d A_{\m_1 ...\m_4}    &=& \d_F  A_{\m_1 ...\m_4} -
      \tfrac{3i}{8}\e_{\g\d}  A^\g_{[\m_1 \m_2} \d_F A^\d_{\m_2 \m_4 ]}  \quad,
\nonumber\\
   \d A^\a_{\m_1 ...\m_6} &=& \d_F A^\a_{\m_1 ...\m_6} +40 A_{[\m_1 ...\m_4}\d_F A^\a_{\m_5 \m_6 ]}
      -20 \d_F A_{[\m_1 ...\m_4 }  A^\a_{\m_5 \m_6 ]} \quad,
\nonumber\\
   \d A^{\a\b}_{\m_1 ...\m_8} &=&  \d_F A^{\a\b}_{\m_1 ...\m_8}
    + \tfrac{21}{4} A^{(\a}_{[\m_1 ...\m_6} \d_F A^{\b)}_{\m_7 \m_8 ]}
    - \tfrac{7}{4} A^{(\a}_{[\m_1 \m_2}\d_F A^{\b)}_{\m_3 ...\m_8 ]}
  \label{linearsusy} \end{eqnarray}
were derived in \cite{Bergshoeff:2005ac}. They have a
particularly simple form, as pointed out in
\cite{Bergshoeff:2006qw}, in which all terms are at most linear in
the gauge fields. Here we denote with $\d_F$ the part of the
supersymmetry transformation that only involves fermi bilinears,
that are \cite{Bergshoeff:2005ac}
  \begin{eqnarray}
  & & \d_F A^\a_{\m_1 \m_2} = 4 i V^\a_- \bar{\e}^* \g_{[\m_1} \psi_{\m_2 ]} +
  V^\a_- \bar{\e} \g_{\m_1 \m_2} \lambda + {\rm c.c.}\,, \nonumber \\
  & & \d_F A_{\m_1 ...\m_4} = \bar{\e} \g_{[\m_1 ...\m_3} \psi_{\m_4 ]} + {\rm c.c.}\,,
  \nonumber \\
  & & \d_F A^\a_{\m_1 ...\m_6} =  12 V^\a_- \bar{\e}^* \g_{[\m_1 ...\m_5} \psi_{\m_6 ]} +
  i V^\a_- \bar{\e} \g_{\m_1 ... \m_6} \lambda + {\rm c.c.}\,, \nonumber \\
  & & \d_F A^{\a\b}_{\m_1 ...\m_8} = 8 V^{( \a}_+ V^{\b )}_- \bar{\e} \g_{[\m_1 ...\m_7} \psi_{\m_8 ]} +
  i V^\a_- V^\b_- \bar{\e}^* \g_{\m_1 ...\m_8} \lambda + {\rm c.c.} \quad
  , \label{purelyfermionicIIB}
  \end{eqnarray}
where all conventions are as in \cite{Schwarz:1983qr}.

The commutators of two supersymmetry transformations on the fields
and dual fields of type IIB were analysed in
\cite{Bergshoeff:2005ac} at lowest order in the fermions. Given the
transformations of eq. \eqref{linearsusy}, together with the
transformations of the scalars and the vielbein
  \begin{eqnarray}
  & & \d V^\a_+ = V^\a_- \bar{\e}^* \l\,, \nonumber \\
  & & \d e_\m{}^a = i \bar{\e} \g^a \psi_\m + {\rm c.c.}
  \end{eqnarray}
and the transformations of the fermions (without including cubic
fermi terms)
  \begin{eqnarray}
    & & \d \psi_\m = D_\m \e +\tfrac{i}{480} F_{\m\n_1 ...\n_4 } \g^{\n_1 ...\n_4 } \e
  +\tfrac{1}{96} G^{\n\r\s} \g_{\m\n\r\s} \e^*
  -\tfrac{3}{32}
  G_{\m\n\r}^{} \g^{\n\r} \e^* \quad ,
\nonumber \\
  & & \d \lambda = i P_\m \g^\m \e^* -\tfrac{i}{24}
  G_{\m\n\r}^{} \g^{\m\n\r} \e \quad , \label{fermisusylowestIIB}
 \end{eqnarray}
where
    \begin{equation}
  P_\m = -\e_{\a\b}  V_+^\a
  \de_\m V_+^\b
  \end{equation}
and
    \begin{equation}
  G_{\m\n\r}^{} = - \e_{\a\b} V^\a_+ F^\b_{\m\n\r} \quad ,
  \end{equation}
the commutators of two supersymmetry transformations on the bosons
close on all the local symmetries of the theory, including the gauge
transformations of eq.~\eqref{bosgaugeIIB}, provided that the duality
relations
  \begin{eqnarray}
  & & F^\a_{\m_1 ...\m_7} = -\tfrac{i}{3} \e_{\m_1 ...\m_7 \n_1
  ...\n_ 3} V^{(\a}_+ V^{\b )}_- \e_{\b\g} F^{\g , \n_1 ...\n_3}\,,
  \nonumber \\
  & &
  F^{\a\b}_{\m_1 \dots \m_9} = i \e_{\m_1 \dots \m_9}{}^\s [
  V^\a_+ V^\b_+ P^*_\s - V^\a_- V^\b_- P_\s ]
  \label{duality3791lowest}
  \end{eqnarray}
hold, together with the self-duality condition for the 5-form
field-strength. What will be crucial in the following are the
expressions for the gauge parameters of the gauge transformations
resulting from the commutators of two supersymmetry transformations
that are purely fermi bilinears, that are
    \begin{eqnarray}
  & & \Lambda_\mu^\a = -2i  V^\a_- \bar{\e}_2^* \gamma_\m \epsilon_{1} +
  {\rm c.c.}\,,
  \nonumber \\
  & &
 \Lambda_{\m\n\r} = -\tfrac{1}{4}  \bar{\e}_2
 \gamma_{\m\n\r} \epsilon_1 +
  {\rm c.c.}\,,
  \nonumber \\
  & & \Lambda_{\m_1 ...\m_5}^\a = - 2 V^\a_- \bar{\e}_2^* \gamma_{\m_1 ...\m_5} \epsilon_{1 }
  + {\rm c.c.}\,,
  \nonumber \\
  & & \Lambda_{\m_1 ...\m_7}^{\a\b} = - V^{(\a}_+ V^{\b )}_- \bar{\e}_2 \gamma_{\m_1 ...\m_7} \epsilon_1
  + {\rm c.c.} \quad .
  \end{eqnarray}

In \cite{Schwarz:1983qr} the closure of the supersymmetry algebra on
the scalars, the vielbein, the 2-forms and the 4-form, as well as on
the fermions, was obtained at all orders in the fermions. Given that
the supersymmetry algebra closes on-shell, this analysis was used to
derive the field equations requiring the closure of the algebra on
the fermi fields. Here we want to perform a similar analysis for all
the bosonic fields and their duals. For simplicity we will only
consider terms that are quadratic in the gravitino, that is we will
ignore all the higher order fermi terms containing the spinor
$\lambda$. The advantage of this is that the modification of the
supersymmetry transformations of eq. \eqref{fermisusylowestIIB} and
of the duality relations of eq. \eqref{duality3791lowest} are all
determined by supercovariance as far as these terms are concerned.
As it turns out, this analysis is sufficient to determine all the
10-forms that are compatible with supersymmetry, as will be shown in
the next subsection.

The expressions for the supercovariant spin connection and field
strengths (only considering terms quadratic in the gravitino) are
  \begin{eqnarray}
  & & \hat{\omega}_{\m ab} = \omega_{\m ab} + i e^\n{}_a e^\r{}_b [
  \bar{\psi}_\m \g_{[\n}\psi_{\r ]} + \bar{\psi}_{[\n} \g_{\r ]} \psi_\m
  + \bar{\psi}_{[ \n} \g_{| \m |}\psi_{\r ]} ]\,, \nonumber \\
  & & \hat{F}_{\m_1 ...\m_3}^\a = F_{\m_1 ...\m_3}^\a + [ -6i V^\a_-
  \bar{\psi}^*_{[\m_1} \g_{\m_2 } \psi_{\m_3 ]} +{\rm c.c.} ]\,,
  \nonumber \\
  & & \hat{F}_{\m_1 ...\m_5} = F_{\m_1 ...\m_5} -5 \bar{\psi}_{[\m_1
  } \g_{\m_2 ...\m_4} \psi_{\m_5 ]}\,, \nonumber \\
  & & \hat{F}^\a_{\m_1 ...\m_7} = F^\a_{\m_1 ...\m_7} + [ -42 V^\a_-
  \bar{\psi}^*_{[\m_1} \g_{\m_2 ...\m_6 } \psi_{\m_7 ]} + {\rm c.c.} ]\,,\nonumber \\
  & & \hat{F}^{\a\b}_{\m_1 ...\m_9} = F^{\a\b}_{\m_1 ...\m_9}  -72
  V^{(\a}_+ V^{\b)}_-
  \bar{\psi}_{[\m_1} \g_{\m_2 ...\m_8 } \psi_{\m_9 ]} \quad .
  \label{supercovIIB}
   \end{eqnarray}

The terms of the form $\epsilon^2 \psi^2$ resulting in the
commutators of two supersymmetry transformations on the form fields
have two sources. The first are the terms, that we schematically
write as $[\d_F , \d_F ]A$, resulting from considering only the
purely fermionic term in the supersymmetry variation of the form
field, that is only the first term on the right hand side of each
line of eq. \eqref{linearsusy}. The resulting $\epsilon^2 \psi^2$
terms can be immediately read by simply substituting the
supercovariant quantities of eq. \eqref{supercovIIB} to the bosonic
result. The second source comes from the purely fermionic variation
of the form fields in the $A \d_F A$ terms in eq.
\eqref{linearsusy}, that is the terms $\d_F A \d_F A$. These can be
immediately written using eq. \eqref{purelyfermionicIIB}, and in
order to compare them to the previous ones one has to perform some
Fierz rearrangements, using the Fierz identity
  \begin{equation}
  \xi \bar{\chi} = -\frac{1}{16} \g_\m (\bar{\chi} \g^\m \xi )
  +\frac{1}{96} \g_{\m\n\r} (\bar{\chi} \g^{\m\n\r} \xi ) -
  \frac{1}{3840} \g_{\m\n\r\s\t} (\bar{\chi} \g^{\m\n\r\s\t} \xi )
  \quad .
  \label{fierzsamechirality}
  \end{equation}
Here $\chi$ and $\psi$ are two generic ten-dimensional spinors
of the same chirality.

The final result is that the commutator of two supersymmetry
transformations on the bosons generates a supersymmetry
transformation of parameter
  \begin{equation}
  \zeta = - \xi^\m \psi_\m \quad , \label{susyparameterIIB}
  \end{equation}
where
  \begin{equation}
  \xi_\m = i \bar{\e}_2 \g_\m \e_1 +{\rm c.c.} \quad .
  \end{equation}
This is the supersymmetry parameter of \cite{Schwarz:1983qr} as far
as  the gravitino terms are concerned.

\subsection{Ten-form Potentials}
We now want to extend this analysis to the 10-forms. In
\cite{Bergshoeff:2005ac} it was shown that the supersymmetry algebra
closes at lowest order in the fermions of a quadruplet and a doublet
of 10-forms whose supersymmetry transformations are
  \begin{eqnarray}
    & &  \d A^{\a\b\g}_{\m_1 ...\m_{10}} = - \tfrac{20}{3} V^{( \a}_+ V^\b_-
    V^{\g )}_- \bar{\e}^* \g_{[\m_1 ...\m_9} \psi_{\m_{10}]} - i V^{( \a}_+ V^\b_-
    V^{\g )}_- \bar{\e} \g_{\m_1 ...\m_{10}} \lambda + {\rm
    c.c.}\nonumber \\
    & & \quad \qquad
    -12  A^{(\a\b}_{[\m_1 ...\m_8}\d_F A^{\g)}_{\m_9 \m_{10}]}
    +3 A^{(\a}_{[\m_1 \m_2} \d_F A^{\b\g)}_{\m_3 ...\m_{10}]}\,, \nonumber \\
   & & \d A^\a_{\m_1 ...\m_{10}} = 20 i V^\a_- \bar{\e}^* \g_{[\m_1 ...\m_9} \psi_{\m_{10}]}
   + V^\a_- \bar{\e} \g_{\m_1 ...\m_{10}} \lambda + {\rm c.c.}\,,
   \label{susytransfquadruplet}
   \end{eqnarray}
where the quadruplet has a non-trivial gauge transformation
   \begin{equation}
\d A^{\a\b\g}_{\m_1 ...\m_{10}} = 10 \de_{[\m_1}
\Lambda^{(\a\b\g)}_{\m_2 ...\m_{10}]} -\tfrac{2}{3} F^{(\a\b}_{[\m_1
...\m_9} \Lambda^{\g)}_{\m_{10}]}
  + 32 F^{(\a}_{[\m_1 ...\m_3} \Lambda_{\m_4 ...\m_{10}]}^{\b\g)}
 \end{equation}
 while the gauge transformation of the doublet is trivial:
 \begin{equation}
 \d A_{\m_1 ...\m_{10}}^\a = 10 \de_{[\m_1} \Lambda_{\m_2
 ...\m_{10}]}^\a \quad .
 \end{equation}

It turns out that there is an additional doublet of 10-forms
$\tilde{A}_{\m_1 ...\m_{10}}$ on which the supersymmetry algebra
closes at lowest order in the fermions. The supersymmetry
transformation of this additional 10-form is
  \begin{eqnarray}
  & & \delta \tilde{A}_{10}^\alpha = - V^\a_- \bar{\e} \g_{\m_1 ...\m_{10}} \lambda + {\rm c.c.} + {9} i
  \epsilon_{\beta \gamma} A_{[\m_1 \m_2}^\beta \delta_F A_{\m_3 ...\m_{10}]}^{\gamma \alpha} +
  252 A_{[\m_1 ...\m_4} \delta_F A_{\m_5 ...\m_{10}]}^\alpha \nonumber \\
  & & \quad \qquad - 378 A_{[\m_1 ...\m_6}^\alpha \delta_F A_{\m_7 ...\m_{10}]} +36 i
  \epsilon_{\beta \gamma} A_{[\m_1 ...\m_8}^{\alpha \beta} \delta_F A_{\m_9
  \m_{10}]}^\gamma \label{nontrivialdoubletsusy}
  \end{eqnarray}
while its gauge transformation is
     \begin{eqnarray}
  & & \delta \tilde{A}_{\m_1 ...\m_{10}}^\alpha = 10 \partial_{[\m_1} \Lambda_{\m_2 ...\m_{10}]} + 2i \epsilon_{\beta
  \gamma} \Lambda_{[\m_1}^\beta F_{\m_2 ...\m_{10}]}^{\gamma \alpha} + 144
  \Lambda_{[\m_1 ...\m_3}
  F_{\m_4 ...\m_{10}]}^\alpha \nonumber \\
  &  &\quad \qquad - \tfrac{2268}{ 5} \Lambda_{[\m_1 ...\m_5}^\alpha F_{\m_6 ...\m_{10}]} + 96 i
  \epsilon_{\beta \gamma} F_{[\m_1 ...\m_3}^{\beta} \Lambda_{\m_4 ...\m_{10}]}^{\gamma \alpha}
  \quad . \label{gaugetransfnewdoublet}
  \end{eqnarray}
In order to prove that the commutator of two supersymmetry
transformations of eq.~\eqref{nontrivialdoubletsusy} closes on the gauge transformations of
eq.~\eqref{gaugetransfnewdoublet}
one makes use of the crucial identities
   \begin{eqnarray}
   & &
   F_{[\m_1 ...\m_5} \Lambda_{\m_6 ...\m_{10}]}^\alpha =0\,,
   \nonumber \\
   & &
   i \epsilon_{\b \g} \Lambda^{\a \b}_{[\m_1 ...\m_7} F_{\m_8 ...\m_{10}]}^\g = -2
   \Lambda_{[\m_1 ...\m_3}
   F_{\m_4 ...\m_{10}]}^\alpha\,,
   \nonumber \\
   & &
   2 i V^\alpha_- P_{[\m_1} \bar{\e}_2 \gamma_{\m_2 ...\m_{10}]} \epsilon_{1 }^* + 2i
   V^\alpha_+ P_{[\m_1}^* \bar{\e}_{2 }^* \gamma_{\m_2 ...\m_{10}]} \epsilon_{1}
   = i \epsilon_{\b\g}
   F_{[\m_1 ...\m_9}^{\alpha \beta} \Lambda_{\m_{10}]}^\g \quad ,
   \end{eqnarray}
which are a consequence of the duality relations of eq.
\eqref{duality3791lowest} and of the properties of the gamma
matrices in ten dimensions. The reason why this additional doublet
of 10-forms was missed in \cite{Bergshoeff:2005ac} is because these
identities were not used in those calculations. Of course, the
supersymmetry algebra closes at lowest order in the fermions on any
linear combinations of the trivial and the non-trivial doublet, and
combining the non-trivial doublet with the trivial one does
not change the form of the gauge transformations of eq.~\eqref{gaugetransfnewdoublet}.

We now show that the non-trivial IIB doublet of 10-forms is
precisely the one predicted by $E_{11}$.\,\footnote{We ignore here the ambiguity
related to the fact that one can always add a trivial IIB doublet (times a constant) to a non-trivial IIB doublet.
The same applies to the IIA case.} The $E_{11}$ analysis of
the generators that is relevant for the IIB theory was performed
originally in \cite{Schnakenburg:2001ya}, while all the form
generators were classified in \cite{Kleinschmidt:2003mf}. The
algebra involving all the form generators associated to the
propagating form fields and the quadruplet of 10-form generators was
derived in \cite{West:2005gu}, where it was also shown that the
symmetry of the group element exactly reproduces the gauge
transformations of the corresponding fields as obtained in
\cite{Bergshoeff:2005ac}. Including also the doublet of 10-form
generators this algebra is
  \begin{eqnarray}
  & & [ R^{\m_1 \m_2}_\a , R^{\m_3 \m_4}_\b ] = i \e_{\a\b} R^{\m_1
  ...\m_4} \quad
  [ R^{\m_1 \m_2}_\a , R^{\m_3 ...\m_6 } ] = R^{\m_1 ...\m_6}_\a
  \quad[ R^{\m_1 \m_2}_\a , R^{\m_3 ... \m_8}_{\b} ] = R^{\m_1
  ...\m_8}_{\a\b} \nonumber \\
  & &   [ R^{\m_1 ...\m_4} , R^{\m_5 ...\m_{10}}_\a ] = R^{\m_1
  ...\m_{10}}_\a \quad [ R^{\m_1 \m_2}_\a , R^{\m_3 ... \m_{10}}_{\b\g} ] = R^{\m_1
  ...\m_{10}}_{\a\b\g} + \tfrac{2}{3} i \e_{\a(\b} R^{\m_1
  ...\m_{10}}_{\g )}
  \end{eqnarray}
with all the other commutators vanishing. One then considers the
group element
  \begin{equation}
  g = e^{B_{\m_1 ...\m_{10}}^{\a\b\g} R^{\m_1 ...\m_{10}}_{\a\b\g} }
  e^{B_{\m_1 ...\m_{10}}^{\a} R^{\m_1 ...\m_{10}}_{\a} } ...
  e^{B_{\m_1 \m_2}^\a R^{\m_1 \m_2}_\a }\quad ,
  \end{equation}
where the $B$'s are the fields associated to each generator.
Requiring symmetry under global transformations of the form $g
\rightarrow g_0 g$ gives the global transformations of the fields,
and in particular for the fields up to the 10-forms one gets
  \begin{eqnarray}
  & & \d B_{\m_1 \m_2 }^\a = a_{\m_1\m_2}^\a\,, \nonumber \\
  & & \d B_{\m_1 ...\m_4} = a_{\m_1 ...\m_4} + \tfrac{i}{2} \e_{\a\b} a_{[ \m_1
  \m_2}^\a  B_{\m_3 \m_4 ]}^\b\,, \nonumber \\
  & & \d B_{\m_1 ...\m_6}^\a = a_{\m_1 ...\m_6}^\a + a_{[\m_1 \m_2
  }^\a B_{\m_3 ...\m_6 ]} + \tfrac{i}{6} \e_{\b\g} a_{[\m_1 \m_2}^\b
  B_{\m_3 \m_4 }^\g B_{\m_5 \m_6 ]}^\a\,, \nonumber \\
  & & \d B_{\m_1 ...\m_8}^{\a\b} = a_{\m_1 ...\m_8}^{\a\b} +
  a_{[\m_1 \m_2}^{(\a} B_{\m_3 ...\m_8 ] }^{\b)} + \tfrac{i}{24}
  \e_{\g\d} a_{[\m_1 \m_2}^\g B_{\m_3 \m_4}^\d B_{\m_5 \m_6}^\a
  B_{\m_7 \m_8 ]}^\b \quad . \label{E11IIB}
  \end{eqnarray}
One then recovers the gauge transformations of the fields by
promoting the constant shifts to gauge transformations:
  \begin{equation}
  a_{\m_1 ... \m_n} = n \de_{[\m_1} \Sigma_{\m_2 ... \m_n]}
  \label{globalisgauge}
  \quad .
  \end{equation}
The algebraic construction that in general leads to the gauge
transformations starting from the global $E_{11}$ transformations
was derived in \cite{Riccioni:2009hi}. One can  show that after
field redefinitions and redefinitions of the gauge parameters, the
transformations of eq. \eqref{E11IIB} coincide with those of eq.~\eqref{bosgaugeIIB}.
Similarly, one can determine from $E_{11}$ the
transformation of the 10-form doublet $B_{\m_1 ...\m_{10}}^\a$.
After reinterpreting the global shifts as gauge transformations as
in eq. \eqref{globalisgauge} one obtains
  \begin{eqnarray}
  & & \d B_{\m_1 ...\m_{10}}^\a = 10 \de_{[\m_1} \Sigma_{\m_2
  ...\m_{10}]}^\a + 4 \de_{[\m_1 } \Sigma_{\m_2 ...\m_4} B_{\m_5
  ...\m_{10}]}^\a + \tfrac{4}{3}i \e_{\b\g} \de_{[\m_1}
  \Sigma_{\m_2}^\b B_{\m_3 ...\m_{10}]}^{\g\a} \nonumber \\
  &&  \quad \quad + \de_{[\m_1}
  \Sigma_{\m_2}^\a B_{\m_3...\m_6 } B_{\m_7 ...\m_{10}]} +
  \tfrac{i}{3} \e_{\b\g} \de_{[\m_1} \Sigma_{\m_2}^\b B_{\m_3
  \m_4}^\g B_{\m_5 \m_6}^\a B_{\m_7 ...\m_{10}]} \quad .
  \end{eqnarray}
After field redefinitions and redefinitions of the gauge parameters
one can show that this gauge transformation coincides with the one
in eq. \eqref{gaugetransfnewdoublet}. This thus shows that the new
doublet of 10-forms $\tilde{A}^\a_{\m_1 ...\m_{10}}$ is the one
predicted by $E_{11}$.

We now consider the commutator of two supersymmetry transformations
on the 10-forms of IIB supergravity, only considering the terms that
do not contain the spinor $\lambda$. For the case of the quadruplet,
the result is exactly as for the lower rank forms discussed in the
previous subsection, and the commutator of two supersymmetry
transformations generates a supersymmetry transformation with
parameter as given in eq. \eqref{susyparameterIIB}. The picture changes
when one considers the two doublets. One can immediately show using
the ten-dimensional Fierz identities of eq.
\eqref{fierzsamechirality} that the supersymmetry algebra does {\it
not} close on both the trivial doublet transforming as in eq.
\eqref{susytransfquadruplet} and on the non-trivial doublet
transforming as in eq. \eqref{nontrivialdoubletsusy}. Only for a
particular combination of these two fields one obtains closure, and
the result is that the only doublet of 10-forms compatible with
supersymmetry is
  \begin{equation}
  \tilde{A}_{\m_1 ...\m_{10}}^\alpha -\tfrac{23}{16} A_{\m_1 ...\m_{10}}^{\alpha} \quad
  . \label{thisistheonlydoublet}
  \end{equation}
This analysis thus produces the intriguing result that for
top-forms the closure of the supersymmetry algebra at lowest order
in the fermions does not in general guarantee actual closure at the
full level. As we will see in the next section, the same result
applies to the IIA case.

\subsection{IIB Superspace}
The superspace version of this story is of course equivalent to the
component one just described, but the organisation of the calculation
differs somewhat. In the superspace approach it is preferable to work
with tensorial quantities, rather than gauge potentials, so that
supersymmetry as well as gauge invariance is manifest at every step.
On the other hand, the introduction of field strengths in the
odd (spinorial) directions as well as the even (spacetime) ones, and the
fact that each field is now a superfield, means that constraints must be
imposed in order to get rid of the non-physical fields. The procedure is
therefore to impose these on the various field strengths and then
to check that they are consistent by examining the Bianchi identities.
It is actually rather easy to find the constraints when one knows
the field content of the theory simply by using dimensional analysis.
A feature of this approach is that we can examine the field strength
even for a ten-form potential because an eleven-form need not vanish
in the superspace context due to the fact that the odd basis differential
forms are commutative.

For the IIB case, the full theory was written down in terms of the
usual physical fields in \cite{Howe:1983sra} and then extended to
include the dual forms in \cite{Dall'Agata:1998va}; later, in
\cite{Bergshoeff:2007ma}, all of these plus the eleven-form field
strengths were included. The full list of Bianchi identities and the
non-vanishing components of all of the forms can be found there;
here we shall just re-examine the eleven-forms. There is a
quadruplet $F_{11}^{\a\b\g}$ which obeys the Bianchi \be d
F_{11}^{\a\b\g}= F_{3}^{(\a} F_{9}^{\b\g)}\quad ,
\label{quadbianchi} \ee and a doublet, $F_{11}^{\a}$, for which the
Bianchi identity is \be
 dF_{11}^\a=\frac{4}{23}\left( \e_{\b\g} F_{3}^\b F_{9}^{\g\a}-\frac{3}{4} F_{5}F_{7}^\a
 \right)\,.\
 \label{dblebianchi}
\ee

Any $n$-form in superspace can be split up into a sum of $(p,q)$-forms,
where $p(q)$ denotes the number of even (odd) indices and
where $n=p+q$.\footnote{Note that this splitting is invariant with respect to a class of
preferred non-coordinate basis frames.}
For an $n$-form field strength $F$, the top component, $F_{n,0}$, has
dimension one, so that the only other ones which can be non-zero are
$F_{n-2,2}$ and $F_{n-1,1}$ which have dimensions zero and one-half
respectively. In a $U(1)$ frame (reached by means of the scalar field
matrix $V$ acting on the $SL(2,R)$ indices) the dimension-zero component
will be a gamma-matrix times
some internal invariant if appropriate, while the dimension one-half
component will be proportional to the dilatino. It will be useful to
think of the symmetric $p$-index gamma-matrices as $(p,2)$ forms, written
$\g_{p,2}$, and the product of a gamma-matrix with the fermion as
a $(p,1)$-form, written $(\g\cdot\l)_{p,1}$. For the eleven-forms
the dimension-zero and one-half components are precisely of this type;
the full details can be
found in \cite{Bergshoeff:2007ma}.

Now we ask if there can be a gauge-trivial doublet of eleven-forms, i.e.
an $F_{11}^\a$ satisfying $d F_{11}^\a=0$. The first non-trivial
component of this identity, at dimension zero, can be written
\be
t_0 F^\a_{9,2}=0\quad ,
\label{dimzero}
\ee
where $t_0$ denotes an algebraic operation formed
by contracting the even-vector index of the dimension-zero torsion,
which is proportional to a gamma-matrix regarded as an even-vector-valued
$(0,2)$-form, with one of the even indices of the form being operated on,
and where all the remaining odd indices are symmetrised. It is quite easy to
see that there is no non-trivial gamma-matrix identity that satisfies
\eqref{dimzero}, so that
$F^\a_{9,2}=0$. But then this implies, using the dimension one-half
Bianchi, that $F^\a_{10,1}$ is also zero, and so the whole of $F$ must vanish.

The component results can be recovered from superspace by observing that a
supersymmetry transformation can be regarded as a super-diffeomorphism with
an odd vector field whose leading component (in an odd coordinate expansion) is identified
with the local supersymmetry parameter in spacetime. It is not difficult to show
that the transformation of a $p$-form potential is given by the interior product
of this vector field with the field strength $F_{p+1}$. The $\l$ terms in the variation
come from $F_{p,1}$ while the gravitino terms come from $F_{p,2}$. The latter
arises because one has to go from a preferred basis to a coordinate basis by means of
the supervielbein, one component of which is the gravitino.

\section{The Top-forms of IIA Supergravity} \label{IIA}
In this section we repeat the same analysis for the IIA case. In
\cite{Bergshoeff:2006qw} the supersymmetry transformations for all
the forms of the IIA theory were derived, and the closure of the
supersymmetry algebra was checked at lowest order in the fermions.
This analysis was performed also in the case of non-vanishing Romans
mass, and apart from all the propagating forms, it was also done for
the 9-form potential, whose field strength is dual to the Romans
mass, and for a non-trivial 10-form and a trivial one. In this
section we will reconsider the analysis of the 10-forms, and for
simplicity we will consider the case of vanishing Romans mass. We
will first review the analysis for all the forms up to the 10-forms.
We will then show that an additional non-trivial 10-form can be
included, while the closure of the supersymmetry algebra at all
orders in the fermions selects two 10-forms out of the three that
are a priori compatible with supersymmetry at lowest order. Finally,
we will perform the same analysis in superspace.

\subsection{IIA supergravity}
We follow the notation of \cite{Bergshoeff:2001pv}, which is the one
also used in \cite{Bergshoeff:2006qw}. The supersymmetry
transformations are thus expressed in the string frame, and we  use
the mostly plus signature, as opposed to the one used in the
previous section. We denote with $C$ the RR fields and with $B$ the
NS-NS fields. The RR fields are forms of odd rank, while the NS-NS
fields are the 2-form, the 6-form and the 8-form. With respect to
ref. \cite{Bergshoeff:2006qw}, we perform field redefinitions for
the 6-form and the 8-form, so that their gauge transformations are
in the abelian basis as is the case for all the other fields. The
resulting gauge transformations are
  \begin{eqnarray}
  & & \d C_{\m_1 ...\m_{2n-1}} = (2n-1) \de_{[\m_1 } \Lambda_{\m_2
  ..\m_{2n-1}]} - {2n-1 \choose 3} H_{[\m_1 ...\m_3} \Lambda_{\m_4
  ...\m_{2n-1}]}\,, \nonumber \\
  & & \d B_{\m_1 \m_2} =2 \de_{[\m_1} \Sigma_{\m_2 ]}\,, \nonumber \\
  & & \d B_{\m_1 ...\m_6} = 6 \de_{[\m_1} \Sigma_{\m_2 ...\m_6 ]} -
  \tfrac{15}{2} G_{[\m_1 \m_2}  \Lambda_{\m_3 ...\m_6 ]} +
  \tfrac{15}{2 } G_{[\m_1 ...\m_4} \Lambda_{\m_5 \m_6 ]} -
  \tfrac{1}{6} G_{\m_1 ...\m_6} \Lambda\,, \nonumber \\
  & & \d B_{\m_1 ...\m_8} = 8 \de_{[\m_1} \Sigma_{\m_2 ...\m_8 ]}
  +21 G_{[\m_1 \m_2}  \Lambda_{\m_3 ...\m_8 ]} -35
  G_{[\m_1 ...\m_4} \Lambda_{\m_5 ... \m_8 ]} + 7
  G_{[ \m_1 ...\m_6} \Lambda_{\m_7 \m_8 ]}\nonumber \\
  & & \quad \qquad  + 28 H_{[\m_1 ...\m_3} \Sigma_{\m_4 ...\m_8 ]}  \quad ,
  \label{gaugealgebraIIA}
  \end{eqnarray}
while the corresponding field strengths are
  \begin{eqnarray}
  & & G_{\m_1 ...\m_{2n}} = 2n \de_{[\m_1 } C_{\m_2 ...\m_{2n}]} -
  {2n \choose 3} H_{[\m_1 ...\m_3} C_{\m_4 ...\m_{2n}]}\,, \nonumber \\
  && H_{\m_1 ...\m_3} = 3 \de_{[\m_1} B_{\m_2 \m_3]}\,, \nonumber \\
  & & H_{\m_1 ...\m_7} = 7 \de_{[\m_1} B_{\m_2 ...\m_7 ]} +
  \tfrac{21}{2} G_{[\m_1 \m_2 } C_{\m_3 ...\m_7 ]} - \tfrac{35}{2}
  G_{[\m_1 ...\m_4} C_{\m_5 ...\m_7 ]} + \tfrac{7}{2} G_{[\m_1
  ...\m_6} C_{\m_7 ]}\,, \nonumber \\
  & & H_{\m_1 ...\m_9} = 9 \de_{[\m_1} B_{\m_2 ...\m_9 ]} -27
  G_{[\m_1 \m_2 } C_{\m_3 ...\m_9 ]} + 63
  G_{[\m_1 ...\m_4} C_{\m_5 ...\m_9 ]} -21 G_{[\m_1
  ...\m_6} C_{\m_7... \m_9 ]} \nonumber \\
  & & \quad \qquad + 42 B_{[\m_1 ...\m_6} H_{\m_7...\m_9 ]} \quad .
  \end{eqnarray}
As in the IIB case, in this basis the supersymmetry transformations
have a particularly simple form, in which all terms are at most
linear in the form fields. The result is
  \begin{eqnarray}
  & & \d C_{\m_1 ...\m_{2n-1}} = \d_F C_{\m_1 ...\m_{2n-1}} + {2n-1
  \choose 2} C_{[\m_1 ... \m_{2n-3} } \delta_F B_{\m_{2n-2} \m_{2n-1}]\,,
  }\nonumber \\
  & & \d B_{\m_1 \m_2} = \d_F B_{\m_1 \m_2}\,, \nonumber \\
  & & \d B_{\m_1 ...\m_6} = \d_F B_{\m_1 ...\m_6} + 3 C_{[\m_1
  ...\m_5} \d_F C_{\m_6]} -10 C_{[\m_1 ...\m_3 } \d_F C_{\m_4
  ...\m_6]} + 3 C_{[\m_1} \d_F C_{\m_2 ...\m_6 ]}\,, \nonumber \\
  & & \d B_{\m_1 ...\m_8} = \d_F B_{\m_1 ...\m_8} - 6 C_{[\m_1
  ...\m_7} \d_F C_{\m_8]} +28 C_{[\m_1 ...\m_5 } \d_F C_{\m_6
  ...\m_8 ]} -14 C_{[\m_1 ...\m_3} \d_F C_{\m_4 ...\m_8 ]} \nonumber \\
  & & \quad \qquad -14 B_{[\m_1... \m_6} \d_F B_{\m_7 \m_8 ]}  \quad , \label{susyformsIIA}
  \end{eqnarray}
where as in the previous section we denote with $\d_F$ the part of
the supersymmetry transformation that only involves fermi bilinears,
that is
  \begin{eqnarray}
    & & \d_F C_{\m_1 ...\m_{2n-1}} = -(2 n-1) \bar{\e} \g_{[\m_1
  ...\m_{2n-2} } \g_{11}^n \psi_{\m_{2n-1}]} + \tfrac{1}{2} \bar{\e}
  \g_{11}^n \g_{\m_1 ...\m_{2n-1} } \l\,, \nonumber \\
  & & \d_F B_{\m_1 \m_2} = 2 \bar{\e} \g_{11} \g_{[\m_1}
  \psi_{\m_2]}\,, \nonumber \\
  & & \d_F B_{\m_1 ...\m_6} = 6 e^{- 2 \phi} \bar{\e} \g_{[\m_1
  ...\m_5} \psi_{\m_6 ]} - e^{-2 \phi} \bar{\e} \g_{\m_1 ...\m_6} \l
  \nonumber\,, \\
& & \d_F B_{\m_1 ...\m_8} = \tfrac{1}{2} e^{-2\phi} \bar{\e}
\g_{\m_1 ...\m_8 } \g_{11} \l \quad
  . \label{purelyfermionicIIA}
  \end{eqnarray}

Given the supersymmetry transformations of the form fields of eq.
\eqref{susyformsIIA}, together with the supersymmetry
transformations of the vielbein and the dilaton
  \begin{eqnarray}
  & & \d e_\m{}^a = \bar{\e} \g^a \psi_\m\,, \nonumber \\
  & & \d \phi = \tfrac{1}{2} \bar{\e} \l\,,
  \end{eqnarray}
as well as the supersymmetry transformations of the fermions at
lowest order in the fermions,
  \begin{eqnarray}
  & &  \d \psi_\m = D_\m \e + \tfrac{1}{8} H_{\m\n\r} \G^{\n\r}
  \G_{11} \e + \tfrac{1}{16} e^{\phi} G_{\n\r} \G^{\n\r} \G_\m \G_{11}
  \e + \tfrac{1}{8 \cdot 4!} e^{\phi} G_{\m_1 \dots \m_4} \G^{\m_1
  \dots \m_4} \G_\m \e\,, \nonumber \\
  & & \d \l = \de_\m \phi \G^\m \e - \tfrac{1}{12} H_{\m\n\r} \G_{11}
  \G^{\m\n\r} \e + \tfrac{3}{8} e^{\phi} G_{\m\n} \G_{11} \G^{\m\n}
  \e + \tfrac{1}{4 \cdot 4!} e^{\phi} G_{\m_1 \dots \m_4} \G^{\m_1
  \dots \m_4} \e \ \ , \label{fermivariationIIA}
  \end{eqnarray}
it was shown in \cite{Bergshoeff:2006qw} that the supersymmetry
algebra closes at lowest order in the fermi fields, provided that
the following duality relations hold:
  \begin{eqnarray}
  & & G_{\m_1 \dots \m_{2n}} = (-1)^n \tfrac{1}{(10-2n)!}
  \e_{\m_1 \dots \m_{2n}}{}^{ \m_{2n+1}\ldots \m_{10}}
  G_{\m_{2n+1} \dots \m_{10}}\,, \nonumber \\
  & &   H_{\m_1 \dots \m_7 } = \tfrac{1}{6} e^{-2\phi} \e_{\m_1 \dots \m_7
  \m\n\r} H^{\m\n\r}\,,\nonumber \\
  & &   H_{\m_1 \dots \m_9} = e^{-2\phi} \e_{\m_1 \dots \m_9 \r} \de^\r
  \phi\,. \label{dualityIIA}
  \end{eqnarray}
The fact that we are considering the massless theory in this
paper implies in particular that $G_{10}$ vanishes as can be seen
from the first equation. The closure of the supersymmetry algebra
implies in particular that the commutator of two supersymmetry
transformations produces the gauge transformations of eq.
\eqref{gaugealgebraIIA}. What will be needed in the following is
the explicit expression for the purely fermionic parts of the
corresponding gauge parameters. These are
  \begin{eqnarray}
  & & \Lambda_{\m_1 ...\m_{2n}} = - e^{-\phi} \bar{\e}_2 \g_{\m_1
  ...\m_{2n}}\g_{11}^{n+1} \e_1\,, \nonumber \\
  & & \Sigma_\m = - \bar{\e}_2 \g_{11} \g_\m \e_1\,, \nonumber \\
  & & \Sigma_{\m_1 ...\m_5} = - e^{-2\phi} \bar{\e}_2 \g_{\m_1
  ...\m_5 } \e_1 \quad \,. \label{susygaugeparametersIIA}
  \end{eqnarray}
Note in particular that there is no purely fermionic
part in the gauge parameter of the 8-form potential.

The analysis of \cite{Bergshoeff:2006qw} can be extended to include
the quartic fermi terms. In particular, if one restricts one's
attention to all terms that do not contain the spinor $\lambda$,
then the modification of the supersymmetry transformations of the
fermions in eq. \eqref{fermivariationIIA} and of the duality
relations of eq. \eqref{dualityIIA} are fully determined by
supercovariance. We thus replace in such equations the spin
connection and the field strengths with the supercovariant
expressions (again neglecting $\lambda$
contributions)\,\footnote{Note that the super-covariant curvature
$\hat{H}_{\m_1 ...\m_9}$ does not contain any gravitino squared
terms.}
\begin{eqnarray}
  & & \hat{\omega}_{\m, ab} = \omega_{\m, ab} + \tfrac{1}{2}
  e^\n{}_a e^\r{}_b [ \bar{\psi}_\n \g_\r \psi_\m + \bar{\psi}_\m \g_\n
  \psi_\r + \bar{\psi}_\n \g_\m \psi_\r ]\,, \nonumber \\
  & & \hat{G}_{\m_1 ...\m_{2n}} = G_{\m_1 ...\m_{2n}} + n (2n-1) e^{-\phi}
  \bar{\psi}_{[\m_1} \g_{\m_2 ...\m_{2n-1} } \g_{11}^n
  \psi_{\m_{2n}]}\,, \nonumber \\
  & & \hat{H}_{\m_1 ...\m_3} = H_{\m_1 ...\m_3} -3
  \bar{\psi}_{[\m_1} \g_{11} \g_{\m_2} \psi_{\m_3 ]}\,, \nonumber \\
  & & \hat{H}_{\m_1 ...\m_7} = H_{\m_1 ...\m_7} -21
  \bar{\psi}_{[\m_1 } \g_{\m_2 ...\m_6} \psi_{\m_7 ]} \quad .
  \label{supercovariantIIA}
  \end{eqnarray}
The calculation then proceeds exactly as in the IIB case discussed
in the previous section. The terms of the form $\epsilon^2 \psi^2$
resulting in the commutators of two supersymmetry transformations on
the form fields are the terms resulting from considering only the
purely fermionic term in the supersymmetry variation of the form
field, that is only the first term on the right hand side of each
line of eq. \eqref{susyformsIIA}, and the terms coming from the
purely fermionic variation of the form fields in eq.
\eqref{susyformsIIA}. The first can be immediately written by simply
substituting the supercovariant expressions of eq.
\eqref{supercovariantIIA} to the bosonic result, while the latter
are simply read from eq. \eqref{purelyfermionicIIA}. In order to
compare the terms, we have to perform some Fierz rearrangements.
Given that the IIA spinors are not chiral, we have to use the Fierz
identity
  \begin{equation}
  \xi \bar{\chi} = -\frac{1}{16} (\bar{\chi} \xi )
  +\frac{1}{32} \g_{\m\n} (\bar{\chi} \g^{\m\n} \xi ) - \frac{1}{384}
  \g_{\m\n\r\s} (\bar{\chi} \g^{\m\n\r\s} \xi ) \quad ,
  \label{fierzoppositechirality}
  \end{equation}
where $\chi$ and $\xi$ are generic ten-dimensional spinors with
opposite chirality, together with the Fierz identity of eq.
\eqref{fierzsamechirality}, which applies when the chirality of the
two spinors is the same. One can then show that the commutator of
two supersymmetry transformations produces a supersymmetry
transformation with parameter
  \begin{equation}
  \zeta = - \xi^\m \psi_\m \quad ,
  \end{equation}
where we denote with $\xi_\m$ the parameter of general coordinate
transformations
  \begin{equation}
  \xi_\m = \bar{\e}_2 \g_\m \e_1 \quad .
  \end{equation}
We now want to repeat this analysis for the 10-forms.

\subsection{Ten-form Potentials}
Using the duality relations of eq. \eqref{dualityIIA} and the
expressions of eq. \eqref{susygaugeparametersIIA}, one derives the
following crucial identities:
  \begin{eqnarray}
  & & \Lambda_{[\m_1 \m_2} G_{\m_3 ...\m_{10}]} = \Lambda_{[\m_1
  ...\m_8} G_{\m_9 \m_{10}]}\,, \nonumber \\
  &  & \Lambda_{[\m_1 ...\m_4} G_{\m_5 ...\m_{10}]} = \Lambda_{[\m_1
  ...\m_6} G_{\m_7 ... \m_{10}]}\,, \nonumber \\
  & & \Sigma_{[\m_1} H_{\m_2 ...\m_{10}]} = \de_{[\m_1} \phi
  e^{-2\phi} \bar{\e}_2 \g_{\m_2 ...\m_{10}]} \e_1 \quad
  .\label{identitiesIIA}
  \end{eqnarray}
Using these identities one can show that the supersymmetry algebra
at lowest order in the fermions closes on {\it two} independent
10-forms transforming non-trivially under gauge transformations. The
gauge transformations of these 10-forms can be written in the
abelian base exactly as for the forms of lower rank. They read
     \begin{eqnarray}
   & & \d B_{\m_1 ...\m_{10}} = 10 \de_{[\m_1} \Sigma_{\m_2 ...\m_{10}]} + \tfrac{135}{2}G_{[\m_1 \m_2}
   \Lambda_{\m_3 ...\m_{10}]} -
   210
   G_{[\m_1 ...\m_6} \Lambda_{\m_7 ...\m_{10}]}\nonumber \\
   & & \quad \qquad  + \tfrac{135}{2} G_{[\m_1 ...\m_8} \Lambda_{\m_9 \m_{10}]}
   - \tfrac{3}{2} G_{\m_1 ...\m_{10}} \Lambda -
   240 H_{[\m_1 ...\m_3} \Sigma_{\m_4 ...\m_{10}]}\,, \nonumber \\
   & & \d \tilde{B}_{\m_1...\m_{10}} = 10 \de_{[\m_1} \tilde{\Sigma}_{\m_2...\m_{10}]} + 315 G_{[\m_1 ...\m_4}
   \Lambda_{\m_5...\m_{10}]}
   - 525 G_{[\m_1 ...\m_6} \Lambda_{\m_7 ...\m_{10}]}\nonumber \\
   & & \quad \qquad
   + 135 G_{[\m_1 ...\m_8} \Lambda_{\m_9 \m_{10}]} -
   3 G_{\m_1 ...\m_{10}} \Lambda - 240 H_{[\m_1 ...\m_3} \Sigma_{\m_4 ...\m_{10}]} \quad
   , \label{twonontrivialIIA}
   \end{eqnarray}
while the supersymmetry transformations in this base are
   \begin{eqnarray}
   & & \d B_{\m_1 ...\m_{10}} = \bar{\e} \g_{\m_1 ...\m_{10}} \l
   - 15 C_{[\m_1 ...\m_9} \d_F C_{\m_{10}]} +
   252 C_{[\m_1 ...\m_5} \d_F C_{\m_6 ...\m_{10}]} \nonumber \\
   & & \quad \quad - 180 C_{[\m_1 ...\m_3} \d_F C_{\m_4 ...\m_{10}]} +
   15 C_{[\m_1} \d_F C_{\m_2 ...\m_{10}]} + 90 B_{[\m_1 ...\m_8} \d_F B_{\m_9 \m_{10}]}\,, \nonumber \\
   & & \d \tilde{B}_{\m_1...\m_{10}} = \bar{\e} \g_{\m_1 ...\m_{10}} \l - 180 C_{[\m_1 ...\m_7} \d_F
   C_{\m_8 ... \m_{10}]} + 630 C_{[\m_1 ...\m_5} \d_F C_{\m_6 ... \m_{10}]} \nonumber \\
   & & \quad \quad -360 C_{[\m_1 ...\m_3} \d_F C_{\m_4 ...\m_{10}]}
   + 30 C_{[\m_1}
   \d_F C_{\m_2...\m_{10}]} + 90 B_{[\m_1 ...\m_8} \d_F B_{\m_9 \m_{10}]} \quad
   . \label{susytransfs10formsIIAnontrivial}
   \end{eqnarray}
This analysis thus completes and corrects the one of ref.
\cite{Bergshoeff:2006qw}, were only one combination of these two
10-forms was found because the identities of eq.
\eqref{identitiesIIA} were basically missed. As shown in
\cite{Bergshoeff:2001pv}, the supersymmetry algebra at lowest order
in the fermions also closes on the trivial 10-form $D_{10}$, whose
supersymmetry transformations is
  \begin{equation}
  \d D_{\m_1 ...\m_{10}} = e^{-2\phi} [ -10 \bar{\e} \g_{[\m_1
  ...\m_9} \psi_{\m_{10}]} + \bar{\e} \g_{\m_1 ...\m_{10}} \l ]
  \quad ,\label{susytrivialIIA}
  \end{equation}
and whose gauge transformation is simply $\d D_{10} = d \Lambda_9$.

Before analysing the supersymmetry algebra on these 10-forms at
quartic order in the fermions, we want to show that the two
non-trivial 10-forms are precisely those predicted by $E_{11}$. The
way to obtain the IIA theory from $E_{11}$ was discussed originally
in \cite{West:2001as}.  The analysis of all the commutation
relations involving the generators up to the 10-form generators, as
well as the computation of all the gauge transformations and the
field strengths for all the fields up to the 10-forms, was performed
in \cite{Riccioni:2009hi}. We refer to eq. (5.1) of that paper
for the algebraic conventions. We add to those commutators the ones
that produce the 10-form generators, which are
  \begin{eqnarray}
  & & [ R^{\m_1 \m_2} , R^{\m_3 ...\m_{10}}] = R^{\m_1 ...\m_{10}}
  \quad \ \   [ R^{\m_1 ...\m_3} , R^{\m_4 ...\m_{10}}] = R^{\m_1
  ...\m_{10}} + 2 \tilde{R}^{\m_1
  ...\m_{10}}\,, \nonumber \\
  & & [ R^{\m_1 ...\m_5} , R^{\m_6 ...\m_{10} } ] =
  \tilde{R}^{\m_1 ...\m_{10}}
  \quad [ R^{\m_1} , R^{\m_2 ...\m_{10}} ] = 4 R^{\m_1
  ...\m_{10}} + 2 \tilde{R}^{\m_1
  ...\m_{10}} \quad ,
  \end{eqnarray}
where $R^{\m_1 ... \m_{10}}$ and $\tilde{R}^{\m_1 ... \m_{10}}$ are
the two independent 10-form generators. If one then considers the
group element
  \begin{equation}
  g = e^{A_{\m_1 ...\m_{10}} R^{\m_1 ...\m_{10}} } e^{\tilde{A}_{\m_1 ...\m_{10}} \tilde{R}^{\m_1 ...\m_{10}} }
  e^{A_{\m_1 ...\m_{9}} R^{\m_1 ...\m_{9}} } ... e^{A_{\m} R^{\m}}\quad ,
  \end{equation}
where the $A$'s are the fields associated to each generator, and
requires symmetry under global transformations of the form $g
\rightarrow g_0 g$, one obtains the global transformations of the
fields. In particular for the 10-forms one gets
  \begin{eqnarray}
  & & \d A_{\m_1 ...\m_{10}} = a_{\m_1 ...\m_{10}} + a_{[ \m_1
  ...\m_3} A_{\m_4 ...\m_{10}]} + a_{[\m_1 \m_2} A_{\m_3
  ...\m_{10}]} + 4 a_{[\m_1} A_{\m_2 ...\m_{10}]} \nonumber \\
  & & \quad -\tfrac{1}{3}
  A_{[\m_1 ...\m_3} A_{\m_4 \m_5 } A_{\m_6 \m_7} A_{\m_8 \m_9}
  a_{\m_{10}]}\,, \nonumber \\
  & & \d \tilde{A}_{\m_1 ...\m_{10}} = \tilde{a}_{\m_1 ...\m_{10}}
  + \tfrac{1}{2} a_{[\m_1 ...\m_5} A_{\m_6 ...\m_{10}]} + 2 a_{[\m_1
  ...\m_3} A_{\m_4 ...\m_{10}]} + a_{[\m_1 \m_2} A_{\m_3 ...\m_5}
  A_{\m_6 ...\m_{10}]} \nonumber \\
  & & \quad + 2 a_{[\m_1} A_{\m_2...\m_{10}]} +
  \tfrac{2}{3} a_{[\m_1} A_{\m_2 ...\m_4} A_{\m_5 \m_6} A_{\m_7
  \m_8} A_{\m_9 \m_{10}]} + \tfrac{1}{2} a_{[\m_1} A_{\m_2 ...\m_6}
  A_{\m_7 \m_8} A_{\m_9 \m_{10}]}\ .\label{E11IIA10forms} \end{eqnarray}
One then recovers the gauge transformations of the fields by
promoting the constant shifts to gauge transformations in a way
analogous to eq. \eqref{globalisgauge}. One can show that after
field redefinitions and redefinitions of the gauge parameters, the
transformations of eq. \eqref{E11IIA10forms} coincide with two
linear combinations of the gauge transformations of eq.
\eqref{twonontrivialIIA}, which shows that the two non-trivial
ten-form we found are exactly those predicted by $E_{11}$.

We now repeat for the IIA 10-forms the same analysis that was
performed for the doublets of 10-forms of the IIB theory. We
consider the commutator of two supersymmetry transformations on the
IIA 10-forms considering all the fermionic terms that are quadratic
in the gravitino. One can immediately see using the ten-dimensional
Fierz identities of eqs. \eqref{fierzsamechirality} and
\eqref{fierzoppositechirality} that the supersymmetry algebra does
{\it not} close on any of the 10-forms transforming under
supersymmetry as in eqs. \eqref{susytransfs10formsIIAnontrivial} and
\eqref{susytrivialIIA}. One only obtains closure by considering a particular combination of each of
the non-trivial 10-forms with the trivial one.
The result is that the only two 10-forms compatible with
supersymmetry are
  \begin{eqnarray}
  & & B_{\m_1 ...\m_{10} } -14 D_{\m_1 ...\m_{10}}\,, \nonumber \\
  & & \tilde{B}_{\m_1 ...\m_{10}} - 38 D_{\m_1 ...\m_{10}}
  \label{thetwoIIAtenforms}
 \quad .
  \end{eqnarray}
This produces for the IIA algebra the same result that we obtained
for IIB in the previous section. The closure of the supersymmetry
algebra on top-forms at lowest order in the fermions is not enough
to guarantee  closure at the full level.

\subsection{IIA Superspace}
The superspace formulation of IIA supergravity was given in
\cite{Carr:1986tk}, while most of the forms were included in
\cite{Cederwall:1996ri,Bergshoeff:1997cf}. It was also derived from
$D=11$ superspace in \cite{Howe:2004ib}, from which paper the
conventions in this subsection are taken. The Bianchi identities for
the RR forms are \be d G_{2n+2}=H_3 G_{2n} \label{RRbianchi} \ee
while for the NS forms one has \bea
dH_3&=&0\,, \nonumber \\
dH_{7}&=& \frac{1}{2} G_4^2 - G_2 G_6\,,\nonumber\\
dH_9&=&-H_3 H_7 +\frac{1}{2} G_4 G_6 -\frac{3}{2} G_2 G_8
\quad ,
\label{NSbianchi}
\eea
The dimension-zero components of the RR field strengths are proportional
to gamma-matrices multiplied by $e^{-\phi}$ in the string frame, while for
the NS field strengths one has no factor of $e^{-\phi}$ in the case of $H_3$,
a factor of $e^{-2\phi}$ for $H_7$, while the dimension-zero component of
$H_9$ vanishes due to the absence of an appropriate symmetric gamma-matrix.
The dimension one-half components depend linearly on $\l$ with the same dilaton factors
($e^{-2\phi}$ for $H_9$).

Now consider the possible eleven-form field strengths. There are two
allowable Bianchi identities that can be combined into one:
\bea
dH_{11}&=&A(H_3 H_9 + \frac{3}{2} G_2 G_{10}-\frac{1}{4} G_6^2) \nonumber\\
&\phantom{=}& +B(-G_2 G_{10} + G_4 G_8 -\frac{1}{2})\quad ,
\label{elevenbianchi}
\eea
where $A$ and $B$ are real constants.
There are also two possible non-trivial dimension zero components, proportional to
$\g_{9,2}$ and $\tilde\g_{9,2}$, where the tilde indicates that a factor of
$\g_{11}$ is present. The second of these requires that both $A$ and $B$ be zero;
we shall come back to this in a moment. For the first case, if we write
\be
H_{9,2}=-iKe^{-2\phi} \g_{9,2}\quad ,
\label{dimzeroa}
\ee
with $K$ constant, we find that \eqref{elevenbianchi} is satisfied if $2A+8B=K$, so that there are
indeed two independent gauge non-trivial eleven-forms. The $(10,1)$ component
of $H_{11}$ is proportional to $e^{-2\phi} (\g\cdot\l)_{10,1}$ multiplied by a constant
depending linearly on $A$ and $B$.

Can there also be a  gauge-trivial eleven-form? The answer is yes,
but that it is itself trivial, i.e. exact. The dimension-zero
component is proportional to $\tilde\g_{9,2}$, and the dimension
one-half component is proportional to $(\tilde\g\cdot\l)_{10,1}$,
but the whole form can be written as $d M_{10}$, where the only
non-zero component of $M$ is $M_{10,0}=\e_{10,0}$ (i.e. $\e$
regarded as a $(10,0)$-form).

\section{Conclusions}\setcounter{equation}{0} \label{conclusions}

In this work we re-considered our earlier work
\cite{Bergshoeff:2006qw,Bergshoeff:2005ac,Bergshoeff:2007ma} on
top-form potentials in IIA and IIB supergravity. We found in both
the IIA and IIB case that the gauge-trivial 10-form potentials found
in our earlier work are excluded by supersymmetry considerations. To
be precise, they are allowed by lowest-order supersymmetry but in
this work we showed that this is not enough. By considering
higher-order fermionic terms we were able to show that gauge-trivial
ten-form potentials are forbidden by supersymmetry. The results of
this paper are confirmed by an independent (IIA as well as IIB)
superspace analysis. Furthermore, all gauge non-trivial top-form
potentials can be derived by a separate $E_{11}$-analysis. This
strongly suggests that we finally obtained full control on the
top-form structure of IIA and IIB supergravity.

It remains an open question what the precise brane interpretation is
of the different gauge non-trivial top-form potentials. It is known
that in the IIB case the D9-brane is part of the quadruplet of
10-form potentials \cite{Bergshoeff:2005ac}. The situation is less
clear for the doublet of eq. \eqref{thisistheonlydoublet} we found
in this work. This doublet does not seem to correspond to a new set
of ``(p,q) 9-branes'' in the usual sense. This can for instance be
seen from the fact that it is impossible to write down a
kappa-symmetric action for a brane which couples to this 10-form
potential. The same applies to the IIA case: the two 10-form
potentials of eq. \eqref{thetwoIIAtenforms} can not lead to a
kappa-symmetric brane effective action. An interpretation of the
10-form potentials as Lagrange multipliers for the constancy of
certain gauge parameter functions $g(x)$ seems also out of the
question  in the absence of any known gauged supergravity in ten
dimensions. A similar lack of interpretation exists in the IIA case.
This is the least to say intriguing given the fact that most (but
not all) other p-forms of IIA and IIB supergravity have a brane
interpretation.

It is natural to ask oneself in which sense the results on the
top-form structure of maximal ten-dimensional supergravity found in
this paper can be extended to other cases with fewer dimensions
and/or supersymmetries. In particular, it would be interesting to
see whether a general pattern emerges and whether this fits with an
extended Kac-Moody algebra structure. These and related questions we
leave for future research.

\subsection*{Acknowledgments}
J.H., T.O. and F.R. wish to thank the University of Groningen for
its hospitality. The work of J.H. was supported by the Swiss
National Science Foundation and the ``Innovations- und
Kooperationsprojekt C-13'' of the Schweizerische
Universit\"atskonferenz SUK/CRUS. The work of T.O.~has been
supported in part by the Spanish Ministry of Science and Education
grant FPA2009-07692, the Comunidad de Madrid grant HEPHACOS
P-ESP-00346 and the Spanish Consolider-Ingenio 2010 program CPAN
CSD2007-00042.  Further, TO wishes to express his gratitude to
M.M.~Fern\'andez for her permanent support. The work of F.R. was
supported by the STFC rolling grant ST/G000/395/1.

\end{document}